\documentclass[twocolumn,aps,prd,preprintnumbers,nofootinbib]{revtex4-2}
\usepackage{graphicx}
\usepackage{amsmath}
\usepackage{amssymb}
\usepackage{amsfonts}
\usepackage{hyperref}
\usepackage{url}
\usepackage{color}
\usepackage{bm}

\usepackage[dvipsnames]{xcolor}

\newcommand{\be}{\begin{equation}}
\newcommand{\ee}{\end{equation}}
\newcommand{\ba}{\begin{eqnarray}}
\newcommand{\ea}{\end{eqnarray}}
\newcommand{\nn}{\nonumber}

\renewcommand{\[}{\begin{equation}}
\renewcommand{\]}{\end{equation}}

\def\lcdm{$\Lambda$CDM }

\usepackage{array}
\def\rat#1{\textcolor{red}{#1}}

\makeatother

\begin{document}

\preprint{IFT-UAM/CSIC-21-57}

\title{Complementary consistency test of the Copernican principle via Noether's theorem and  machine learning forecasts}

\author{Rub\'{e}n Arjona}
\email{ruben.arjona@uam.es}

\author{Savvas Nesseris}
\email{savvas.nesseris@csic.es}

\affiliation{Instituto de F\'isica Te\'orica UAM-CSIC, Universidad Auton\'oma de Madrid,
Cantoblanco, 28049 Madrid, Spain}

\date{\today}

\begin{abstract}
The Copernican principle (CP), i.e. the assumption that we are not privileged observers of the Universe, is a fundamental tenet of the standard cosmological model. A violation of this postulate implies the possibility that the apparent cosmic acceleration could be explained without the need of a cosmological constant, dark energy or paper we present a new test of the CP relating the distance and the expansion rate, derived via Noether's theorem, which is complementary to other tests found in the literature. We also simulate fiducial data based on upcoming stage IV galaxy surveys and use them to reconstruct the Hubble rate $H(z)$ and the angular diameter distance $d_A(z)$ in order to forecast how well our null test can constrain deviations from the cosmological constant model. We find that our new test can easily rule out several scenarios based on the Lema\^{\i}tre-Tolman-Bondi void model at confidence of $\gtrsim 3\sigma$ at middle to high redshifts ($z>0.5$).
\end{abstract}
\maketitle

\section{Introduction}

The standard cosmological paradigm is based on two fundamental assumptions: first, that the dynamics of space-time are governed by Einstein's field equations and second, that the Universe is homogeneous and isotropic at scales larger than $\sim$100 Mpc, a hypothesis normally referred to as the cosmological principle, which is considered to be a generalization of the Copernican principle (CP). The latter is one of the pillars of modern cosmology, stating that we do not occupy a special place in the Universe, or in other words, that any point in space must be equivalent to any other \cite{Clarkson:2007pz}. This leads to the framework of an homogeneous and isotropic background spacetime governed by the Friedmann-Lema\^{\i}tre-Robertson-Walker (FLRW) metric, which describes the geometry of the universe in terms of the scale factor $a(t)$, which obeys the Friedmann equation \cite{Weinberg:1988cp}. 

Clearly, any violations of the CP would disprove homogeneity and would provide an explanation for the observed accelerated expansion of the Universe without the need for a dark energy component. The latter could in fact have several possible explanations, such as modified gravity theories, global inhomogeneities such as a void model or novel dark fluid components currently unobserved in the laboratories \cite{Tomita:2000jj,Barrett:1999fd,Celerier:1999hp}.

Void models have the particularity that they do not employ any form of dark energy component as the accelerating expansion of the Universe is interpreted from the fact that we live close to the center of a large underdense region, thus having the observer in a special place in the local Universe \cite{Clifton:2008hv}, or with backreaction which might mimic acceleration via the nonlinear effect of inhomogeneities  \cite{Buchert:1999er,Wiltshire:2007fg}.

As of now, the CP has been tested with different observations such as radio-astronomy \cite{Bester:2017lrt}, time drift of cosmological redshift \cite{Uzan:2008qp}, using type Ia supernovae 
\cite{Bolejko:2008cm,Clifton:2008hv}, the integrated Sachs Wolfe effect \cite{Tomita:2009wz}, galaxy correlations and the baryon acoustic oscillations \cite{February:2012fp}, the Hubble parameter \cite{Zhang:2012qr}, machine learning and cosmological distance probes \cite{Sapone:2014nna}, peculiar velocities \cite{Hellwing:2016pdl}, the cosmic microwave background (CMB) temperature and polarization spectrum \cite{Zibin:2008vk}, the spectral distortion of the CMB spectrum  \cite{Caldwell:2007yu}, galaxy surveys \cite{Labini:2010qx}, the first
order anisotropic kinetic Sunyaev Zel’dovich (kSZ) effect \cite{Zhang:2010fa,Yoo:2009jba} and finally with a plethora of cosmological data that can be used to constrain spatial homogeneity \cite{Valkenburg:2012td}. Finally, for a representation of the current state of constraints on LTB models using various data see Ref.~\cite{Camarena:2021mjr}.

The simplest inhomogeneous models of the Universe are given by a spherically symmetric distribution of matter, which is mathematically described by
a Lema\^{\i}tre-Tolman-Bondi (LTB) spacetime \cite{Bondi:1947fta}, which has been shown it can produce a Hubble diagram which in the past was consistent with observations, see Ref.~\cite{GarciaBellido:2008nz} for a broad overview and Refs.~\cite{Lukovic:2019ryg,Kazantzidis:2020tko} for some more recent constraints, but with more recent data it has been realized that simple void models cannot be used as an alternative to dark energy. Specifically, LTB models where decaying modes are not present produce a large kSZ signal~\cite{GarciaBellido:2008gd,Zhang:2010fa,Moss:2011ze}, while models with large decaying modes, and correspondingly a small kSZ signal, are not viable due to y-distortions~\cite{Zibin:2011ma}. Specific cases where the void LTB models can be viable require fine-tuned initial conditions, thus leading to questions about the naturalness of these models. 

Since in the near future we are going to have sufficiently good and rich cosmological data we will use the LTB models as a check for our null test to show how they can be ruled out with high confidence.
Probing for deviations from the cosmological constant model ($\Lambda$CDM) is non-trivial in the absence of guiding principles or laboratory data \cite{Shafieloo:2009hi}. Thus, over the years several consistency tests of the \lcdm have appeared in the literature. In general these tests are constructed so that possible deviations from \lcdm at any redshift are apparent and easy to quantify in the form of null tests. These are formulated such that they can be computed using reasonably directly observable quantities at any redshift, thus by computing the consistency test using data from multiple redshifts, one can examine the validity of the basic assumptions of the cosmological standard model. If these assumptions hold, the null test should be independent of redshift. 

In this work we present a complementary test to the well-known curvature test of Ref.~\cite{Clarkson:2007pz} that can be used to falsify the CP. This null test depends solely on distance and Hubble rate observations and is derived with the aid of Noether's theorem. The advantage of our new test is that it does not suffer from divergences and provides tighter constraints at high redshifts, as we will discuss in later sections.

Our paper is organized as follows: In Sec.~\ref{sec:theory} we present the theoretical formalism, our null test named $\mathcal{O}(z)$ and a description of the LTB models used to check the consistency test. In Sec.~\ref{sec:reconstructions} we describe our simulated data based on an optimistic Stage IV galaxy survey and the machine learning (ML) algorithm used to reconstruct the data, namely the genetic algorithms. Finally, in Sec.~\ref{sec:results} we present our results and in Sec.~\ref{sec:conclusions} we summarize our conclusions.

\section{Analysis \label{sec:theory}}
Under the assumption of a Friedmann-Lema\^{\i}tre-Robertson-Walker (FLRW) metric, the luminosity distance can be written as
\begin{equation}\label{eq:dl}
d_{L}(z)=\frac{c(1+z)}{H_{0} \sqrt{-\Omega_{k}}} \sin \left(\sqrt{-\Omega_{k}} \int_{0}^{z} d z^{\prime} \frac{H_{0}}{H\left(z^{\prime}\right)}\right),
\end{equation}
where $\Omega_{k}$ is the curvature parameter today and $H(z)$ is the expansion rate. The luminosity distance $d_L(z)$ is related to the angular diameter distance $d_A(z)$ through the Etherington relation, i.e. $d_L(z)=(1+z)^2d_A(z)$. Using the comoving angular diameter distance $D(z)$, defined as $D(z)=\frac{H_0}{c}(1+z)d_A(z)$, we can regroup Eq.~(\ref{eq:dl}) to solve for the curvature parameter $\Omega_k$ in terms of $H(z)$ and $D(z)$ as \cite{Clarkson:2007pz}
\begin{equation}\label{eq:ok}
\Omega_{k}=\frac{\left[E(z) D^{\prime}(z)\right]^{2}-1}{\left[D(z)\right]^{2}},
\end{equation}
where $E(z)\equiv H(z)/H_0$ is the dimensionless Hubble parameter and the prime is a derivative with respect to the redshift $'=d/dz$. The above relation allows us to estimate the spatial curvature parameter from distance and Hubble rate observations, without having to assume any particular dark energy model or other model parameters. It also allows us to test the curvature at any single redshift as it has been reconstructed in several works, see for instance Refs.~\cite{Clarkson:2007pz,Shafieloo:2009hi,Mortsell:2011yk,Sapone:2014nna,Rasanen:2014mca,LHuillier:2016mtc,Denissenya:2018zcv,Park:2017xbl,Cao:2021ldv,Khadka:2020tlm,Cao:2021ldv}. 
Since the curvature parameter $\Omega_k$ does not depend on redshift, we can differentiate this to obtain a relation that must always equal zero. This can be expressed as \cite{Clarkson:2007pz}
\ba\label{eq:cp}
C(z)&=&1+E(z)^{2}\left(D(z) D^{\prime \prime}(z)-D^{\prime}(z){}^{2}\right)\nn \\
&+&E(z) E^{\prime}(z) D(z) D^{\prime}(z),
\ea
where $C(z)$ has to be zero at all redshifts in any model described by a FLRW metric, as was originally shown in Ref.~\cite{Clarkson:2007pz}. In Sec.~\ref{sec:results} we will present constraints on this test with an upcoming Stage IV survey along with a complementary CP test inspired from Noether's theorem. The advantage of using Noether's theorem to make a complementary test is that by taking into account the symmetries of the system of equations that describe the expansion of the Universe, we can reduce the order of the differential equations that appear in the final test. This allows us to keep the errors of the reconstructions smaller, as higher order derivatives of noise data tend to make the reconstructions less robust at high redshifts, as we demonstrate in what follows. 

\subsection{Lagrangian formalism and null test\label{sec:noether}}
We now present a complementary test of $C(z)$ to probe the CP. Using Eq.~(\ref{eq:ok}) we can solve for $D^{\prime}(z){}^{2}$, which will be given by
\be
D^{\prime}(z){}^{2}=\frac{1+D(z)^2\Omega_k}{E(z)^2},
\ee
and inserting this relation into Eq.~(\ref{eq:cp}) we have
\be\label{eq:cp2}
D^{\prime \prime}(z)+\frac{E^{\prime}(z)}{E(z)}D^{\prime}(z)-\frac{\Omega_k}{E(z)^2}D(z)=0.
\ee
To find a null test that involves the distance measure $D(z)$, we will make use of the Lagrangian formalism. The first step is to find a Lagrangian for Eq.~(\ref{eq:cp2}) and, with the help of Noether’s theorem, to find an associated conserved quantity. For a description of the Noether symmetry approach and applications for null tests see \cite{Nesseris:2014mfa,Nesseris:2014qca}.

In a nutshell, if we assume that the Lagrangian can be written as $\mathcal{L}=\mathcal{L}\left(z, D(z), D^{\prime}(z)\right)$. Then, the
Euler-Lagrange equations are:
\begin{equation}\label{eq:E-L}
\frac{\partial \mathcal{L}}{\partial D}-\frac{d}{d z} \frac{\partial \mathcal{L}}{\partial D^{\prime}}=0.
\end{equation}
So, let us assume a Lagrangian of the form
\ba
\mathcal{L} &=&T-V, \\
T &=&\frac{1}{2} f_{1}(z, E(z))\,D^{\prime}(z)^{2}, \\
V &=&\frac{1}{2} f_{2}(z, E(z))\,D(z)^{2},
\ea
where the $f_1$ and $f_2$ are arbitrary functions that need to be determined so that the resulting equation after implementing the Euler-Lagrange Eq.~(\ref{eq:E-L}) is exactly Eq.~(\ref{eq:cp2}). Therefore, substituting the former Lagrangian in the Euler-Lagrange Eq.~(\ref{eq:E-L}) and comparing the result with Eq.~(\ref{eq:cp2}) we are able to get the two functions $f_1$ and $f_2$ and consequently to build the Lagrangian $\mathcal{L}$ of the system:
\ba\label{eq:lagrange}
\mathcal{L}=\frac{1}{2}E(z)\,D^{\prime}(z){}^{ 2}+\frac{1}{2}\frac{\Omega_k}{E(z)}D(z)^{2}.
\ea
It is easy to see that substituting Eq.~(\ref{eq:lagrange}) into Eq.~(\ref{eq:E-L}) results exactly to Eq.~(\ref{eq:cp2}). Now that we have a Lagrangian we can use Noether’s
theorem to find a conserved quantity that will be later translated to the null test. So, if we have an infinitesimal transformation $\mathbf{X}$ with a generator
\ba 
\mathbf{X} &=&\alpha(D) \frac{\partial}{\partial D}+\frac{d \alpha(D)}{d z} \frac{\partial}{\partial D^{\prime}}, \\ \frac{d \alpha(D)}{d z} & \equiv & \frac{\partial \alpha}{\partial D} D^{\prime}(z)=\alpha^{\prime}(z) ,
\ea
such that for the Lie derivative of the Lagrangian we have $L_{X} \mathcal{L}=0$, then
\begin{equation}\label{eq:sigma}
\Sigma=\alpha(z) \frac{\partial \mathcal{L}}{\partial D^{\prime}},
\end{equation}
is a constant of “motion” for the Lagrangian of Eq.~(\ref{eq:lagrange}). From Eq.~(\ref{eq:sigma}) we get that
\begin{equation}
    \Sigma=\alpha(D(z))E(z)\,D^{\prime}(z),
\end{equation}
while from the Lie derivative we also obtain:
\begin{equation}
\alpha(z)=\alpha_0\,e^{-\int_{0}^{z} \frac{\Omega_k D(x)}{E(x)^2\,D^{\prime}(x)} d x},
\end{equation}
where $\alpha_0$ is an integration constant. Then, the constant $\Sigma$ becomes
\begin{equation}\label{eq:sig2}
  \Sigma=\frac{E(z)D^{\prime}(z)}{D^{\prime}(0)}e^{-\int_{0}^{z} \frac{\Omega_k D(x)}{E(x)^2\,D^{\prime}(x)} d x},
\end{equation}
where we absorbed $\alpha_0$ into $\Sigma$ and normalized the above equation so that the null test must be equal to unity for all values of $z$.

Finally, to write the above null test only as a function of $E(z)$ and $D(z)$ we substitute $\Omega_k$ from Eq.~(\ref{eq:ok}) into Eq.~(\ref{eq:sig2}), then the null test is given by
\begin{equation}\label{eq:Otest}
   \mathcal{O}(z)=\frac{E(z)D^{\prime}(z)}{\, D^{\prime}(0)}e^{-\int_{0}^{z} \frac{E(x)^2\,D^{\prime}(x){}^{2}-1}{E(x^2\,D(x)\,D^{\prime}(x)} d x}.
\end{equation}

\subsection{LTB model}
An alternative explanation, besides the cosmological constant $\Lambda$, for the current phase of accelerated expansion of the Universe is the idea of inhomogeneous universe models, where this expansion can be seen as an effective acceleration induced by our special position as observers residing inside a huge under-dense region of space. These models violate the CP and a simple toy model which has been studied extensively in the literature is the spherically symmetric Lema\^{\i}tre-Tolman-Bondi model \cite{lemaitre1997expanding,Tolman:1934za,Bondi:1947fta} (LTB) which describes a local void. It actually represents a family of models coming from a spherically symmetric solution of Einstein equations exerted by pressureless matter and no cosmological constant, as one still needs to provide a matter density profile \cite{Uzan:2008qp}. The metric for our model of interest is given by
\begin{equation}\label{eq:sa}
d s^{2}=-d t^{2}+X^{2}(r, t) d r^{2}+A^{2}(r, t) d \Omega^{2},
\end{equation}
where $d \Omega^{2}=d \theta^{2}+\sin ^{2} \theta d \phi^{2}$ and the function $A(r, t)$ is analogous to the scale factor of the FLRW metric, albeit it also  has a dependence on both time and the radial coordinate $r$. One can find a relation between $X(r,t)$ and $A(r,t)$ through the $0-r$ component of the Einstein equations, i.e $X(r, t)=A^{\prime}(r, t) / \sqrt{1-k(r)}$, where a prime denotes a derivative with respect to the coordinate $r$ and $k(r)$ represents an arbitrary function, being similar to the role of the spatial curvature parameter. 

This model can be totally described by the matter density $\Omega_m(r)$ and the Hubble expansion rate $H(r)$. We will check our consistency test with a particular LTB model known as the GBH parametrization \cite{GarciaBellido:2008nz}. In this case the matter and Hubble parameter profiles are given by

\ba
\Omega_{\mathrm{m}}(r) &=&\Omega_{\text {out }}+\left(\Omega_{\text {in }}-\Omega_{\text {out }}\right) \frac{1-\tanh \left[\left(r-r_{0}\right) / 2 \Delta r\right]}{1+\tanh \left[r_{0} / 2 \Delta r\right]},\nn \label{eq:GBH1} \\
~ \\
H_{0}(r) &=&H_{0}\left(\frac{1}{\Omega_{\mathrm{k}}(r)}-\frac{\Omega_{\mathrm{m}}(r)}{\sqrt{\Omega_{\mathrm{k}}(r)^{3}}} \sinh ^{-1} \sqrt{\frac{\Omega_{\mathrm{k}}(r)}{\Omega_{\mathrm{m}}(r)}}\right),\label{eq:GBH2}
\ea
where $\Omega_k(r)=1-\Omega_m(r)$, $\Omega_{\text{out}}$ is the value of the matter density at infinity, $\Omega_{\text{in}}$ is the value of the matter density at the center of the void, $r_0$ is the size of the void and $\Delta r$ represents a scale that characterises the transition to uniformity. In Table~\ref{tab:LTB} we show the four GBH parameters used in our analysis, which correspond to characteristic voids of sizes of a few Gpc, as suggested in Ref.~\cite{GarciaBellido:2008nz}.

Note however, that in Ref.~\cite{Camarena:2021mjr} the LTB model, albeit with a different profile than that of Eq.~\eqref{eq:GBH1} and including a cosmological constant, was confronted against a plethora of cosmological data, including the Planck measurements of the CMB spectrum, BAO data, the Pantheon compilation of type Ia supernovae, local H0 measurements, the $H(z)$ cosmic chronometers and finally, the Compton y-distortion
and kinetic Sunyaev–Zeldovich effect. In summary, it was found that the aforementioned data can  tightly constrain the LTB model, almost at the cosmic variance level. In particular, on scales of $\sim 100\,\mathrm{Mpc}$ structures can have a small non-Copernican contrast of just $\delta\sim 0.01$.

\begin{table}[!thb]
    \centering
\begin{tabular}{|cccc|}
\hline$\Omega_{\mathrm{m}, \text{in}}$ & $r[\mathrm{Gpc}]$ & $\Delta r[\mathrm{Gpc}]$ & LTB models \\
\hline \hline \textcolor{red}{0.298} & \textcolor{red}{1.0} & \textcolor{red}{0.30} & \rat{LTB1} \\
\textcolor{PineGreen}{0.197} & \textcolor{PineGreen}{1.5} & \textcolor{PineGreen}{0.45} & \textcolor{PineGreen}{LTB2} \\
\textcolor{blue}{0.156} & \textcolor{blue}{1.8} & \textcolor{blue}{0.54} & \textcolor{blue}{LTB3} \\
\textcolor{Orange}{0.200} & \textcolor{Orange}{2.0} & \textcolor{Orange}{0.60} & \textcolor{Orange}{LTB4}  \\
\hline
\end{tabular}
\caption{The parameters for the LTB models, where in all cases $\Omega_{\mathrm{m}, \text{out}}=1$ and $H_{0}=77$~km/s/Mpc. Note that the actual value of the Hubble rate today as measured by a comoving observer at $z=0$, depends on the specific profile used. Here we assume the constrained GBH LTB profile of Ref.~\cite{GarciaBellido:2008nz}, given by Eqs.~\eqref{eq:GBH1} and \eqref{eq:GBH2}. }
\label{tab:LTB}
\end{table}

\section{Reconstrutions \label{sec:reconstructions}}
We now describe both the mock data used and the machine learning process used to reconstruct the null test, namely the genetic algorithms.

\subsection{Mock data \label{sec:data}}
Our mock baryon acoustic oscillations (BAO) data for the angular diameter distance $d_A(z)$ and the Hubble rate $H(z)$ are based on a future upgrade of Dark Energy Spectroscopic Instrument (DESI) \cite{Aghamousa:2016zmz}. DESI is 
a survey with the goal of probing the expansion rate and large-scale structure (LSS) of the universe and  can complement other future BAO surveys by extending the probed redshift range \cite{Martinelli:2020hud}.

The DESI survey, whose operations started at the end of 2019, is expected to obtain optical spectra for tens of millions of galaxies and quasars up to redshift $z\sim 4$, which will allow for BAO and redshift-space distortion cosmological analyses. Our forecast data will cover the redshift range $z\in [0.05,3.55]$, but their precision will also depend on the target population. The blue galaxies (BGs) will cover the redshift range $z\in [0.05,0.45]$ in five equispaced redshift bins, the luminous red galaxies (LRGs) and emission line galaxies (ELGs) will focus on $z\in [0.65,1.85]$ with $13$ equispaced redshift bins, while the Ly-$\alpha$ forest quasar survey will cover $z\in [1.96,3.55]$ with $11$ equispaced redshift bins \cite{Aghamousa:2016zmz}. The difference in the redshift distributions of the various targets (BGs, ELGs, QSOs, etc) are due to the fact that each of these targets needs different selection methods to accumulate sufficiently large samples of spectroscopic targets from photometric data, see for example Sec. 3 in Ref.~\cite{Aghamousa:2016zmz}.

To create the mocks we assume the $H(z)$ and $d_A(z)$ are uniformly distributed in the range $z \in[0.1,3.55]$, divided into $20$ equally spaced binds of step $dz=0.2$. The $H(z_i)$ and $d_A(z_i)$ function was estimated as its theoretical value from the different cosmological models plus a Gaussian error (which can be either negative or positive) and assigning an error of $0.5\%$ of its value to $H(z)$ and for $d_A(z)$ an error of $0.28\%$ for $z<1.1$ and $0.39\%$ for $z>1.1$, which is in agreement with a similar setup to \cite{Vargas-Magana:2019fvd}. We further assume these measurements to be uncorrelated.

Finally, we have assumed a fiducial cosmology of a flat $\Lambda$CDM model with matter density $\Omega_\mathrm{m,0}=0.3$ and Hubble constant of $H_0=70 \,\mathrm{km}\, \mathrm{s}^{-1}\,\mathrm{Mpc}^{-1}$. 

\subsection{Genetic algorithms}

In this section we explain the implementation of the genetic algorithms (GA) in our analysis. They have successfully been applied in cosmology for several reconstructions on a wide range of data, for further details see Refs.~\cite{Bogdanos:2009ib,Nesseris:2010ep, Nesseris:2012tt,Nesseris:2013bia,Sapone:2014nna,Arjona:2020doi,Arjona:2020kco,Arjona:2019fwb,Arjona:2021hmg,Arjona:2020skf,Arjona:2020axn}. Other applications of the GA cover particle physics \cite{Abel:2018ekz,Allanach:2004my,Akrami:2009hp} and astronomy and astrophysics \cite{wahde2001determination,Rajpaul:2012wu,Ho:2019zap}. Also, other symbolic regression methods applied in physics and cosmology can be found at \cite{Udrescu:2019mnk,Setyawati:2019xzw,vaddireddy2019feature,Liao:2019qoc,Belgacem:2019zzu,Li:2019kdj,Bernardini:2019bmd,Gomez-Valent:2019lny}.

The GA can be seen as a particular type of ML methods mainly constructed to perform unsupervised regression of data, i.e. it carries out non-parametric reconstructions finding an analytic function of one or more variable that describes the data extremely well. 

The GA operates by simulating the notion of biological evolution through the principle of natural selection, as conveyed by the genetic operations of mutation and crossover. Essentially, a set of test functions evolves as time goes by through the effect of the stochastic operators of crossover, i.e the joining of two or more candidate functions to form another one, and mutation, i.e a random alteration of a candidate function. This procedure is then repeated thousands of times so as to ensure convergence, while different random seeds can be used to further explore the functional space.

Due to the construction of the GA as a stochastic approach, the probability that a population of functions will give rise to offspring is normally assumed to be proportional to its fitness to the data, where in our analysis is given by a $\chi^2$ statistic and conveys how good every individual agrees with the data. For the simulated data in our analysis we are assuming that the likelihoods are sufficiently Gaussian that we use the $\chi^2$ in our GA approach. In the GA, the probability to have offspring and the fitness of each individual is proportional to the likelihood, causing an evolutionary pressure that favors the best-fitting functions in every population, hence driving the fit towards the minimum in a few generations.

In our analysis we reconstruct the Hubble rate $H(z)$ and the angular diameter distance $d_A(z)$ from the mock data created, and the course of action to its reconstruction is as follows. First, our predefined grammar was formed on the following  functions: exp, log, polynomials etc. and a set of operations $+,-,\times,\div$, see Table \ref{tab:grammars} for the complete list.

\begin{table}[!thb]
\caption{The grammars used in the GA analysis. Other complex forms are automatically produced by the mutation and crossover operations as described in the text.\label{tab:grammars}}
\begin{centering}
\begin{tabular}{cc}
 Grammar type & Functions \\ \hline
Polynomials & $c$, $x$, $1+x$ \\
Fractions & $\frac{x}{1+x}$\\
Trigonometric & $\sin(x)$, $\cos(x)$, $\tan(x)$\\
Exponentials & $e^x$, $x^x$, $(1+x)^{1+x}$ \\
Logarithms & $\log(x)$, $\log(1+x)$
\end{tabular}
\par
\end{centering}
\end{table}

\begin{figure*}[!thb]
\centering
\includegraphics[width = 0.48\textwidth]{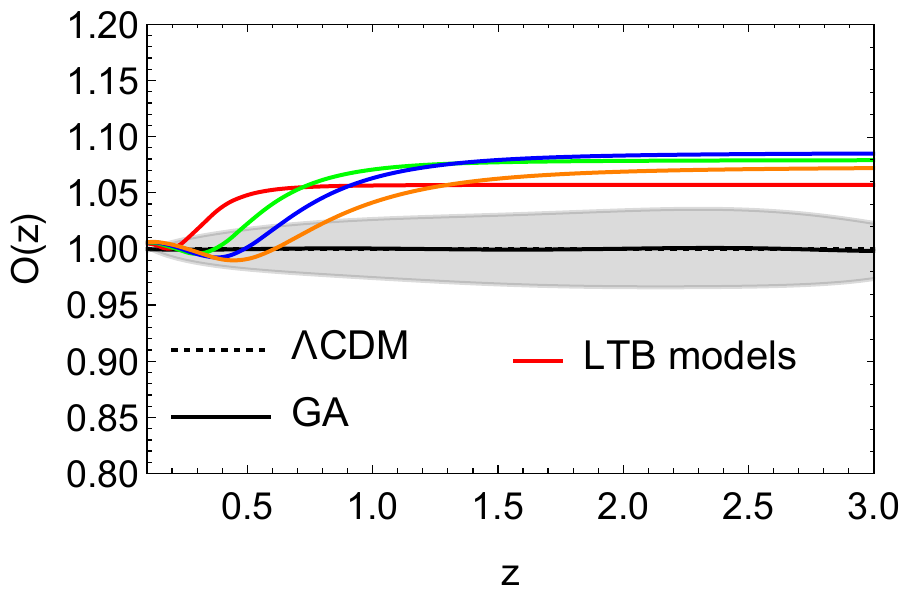}
\includegraphics[width = 0.48\textwidth]{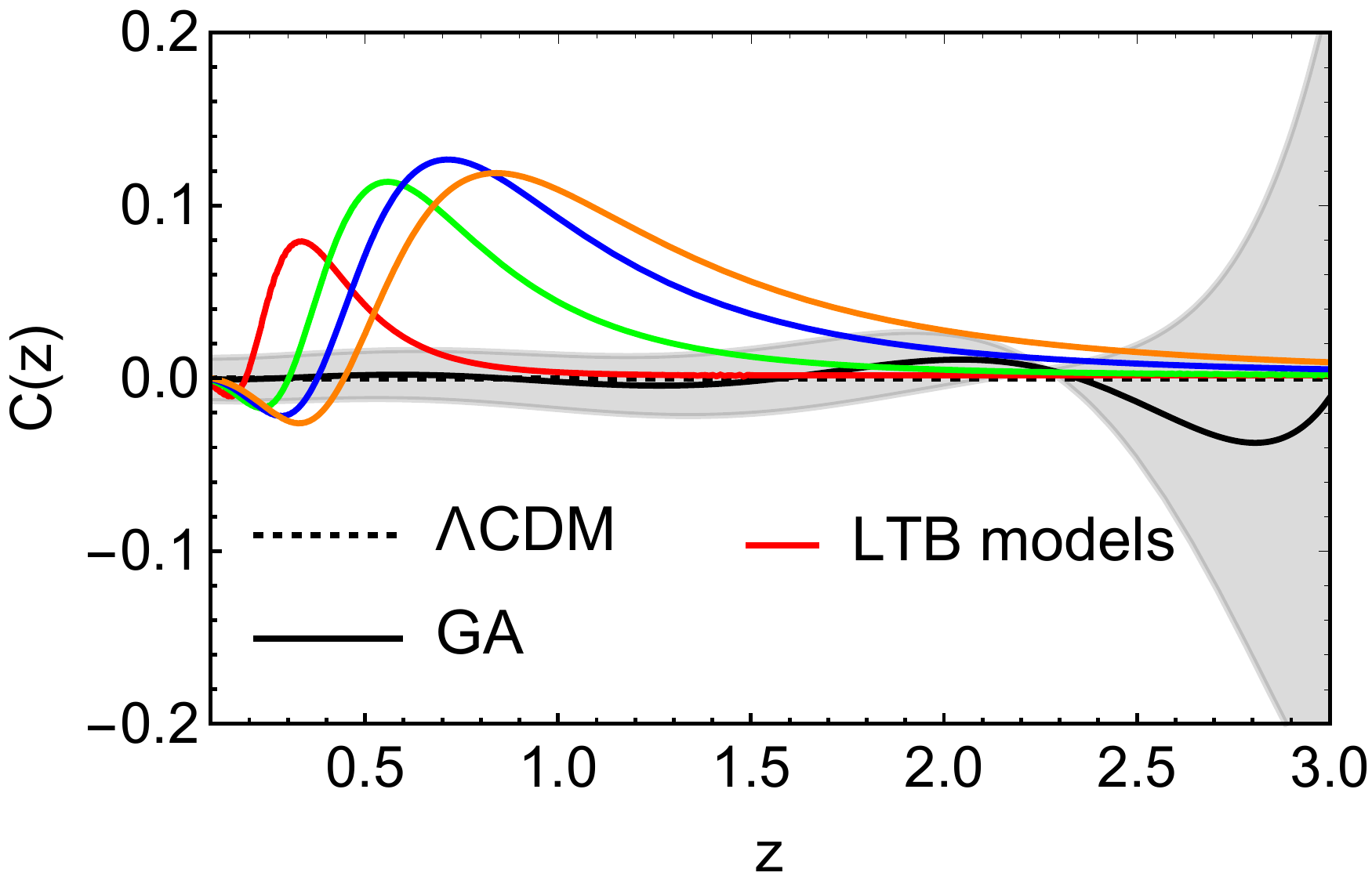}\caption{Reconstructions of the tests of the Copernican principle: the $\mathcal{O}(z)$ test of Eq.~\eqref{eq:Otest} (left panel) and the $C(z)$ test of Eq.~\eqref{eq:cp2} (right panel). The black solid line corresponds to the GA best fit, the gray shaded region corresponds to the $68.3\%$ confidence regions and the colored lines to the various LTB profiles described in Table \ref{tab:LTB}. \label{fig:FS}}
\end{figure*}

As a prior for our $H(z)$ reconstruction we imposed that $H(z=0)=H_0$ and for the $d_A(z)$ reconstruction we assumed that $d_A(z\ll1)=c/H_0\,z+O(z^2)$, where both priors have been motivated by physical reasons, namely the fact that the value of the Hubble parameter today is the one that we actually infer from observations and the Hubble law respectively. However, we make no assumptions on the curvature of the Universe or any modified gravity or dark energy model. Furthermore, in order to avoid overfitting or any spurious reconstructions we  required that all functions reconstructed by the GA are continuous and differentiable, without any singularities in the redshift probed by the data. 

Once the initial population has been constructed, the fitness of each member is computed by a $\chi^2$ statistic, using the $H(z)$ and $d_A(z)$ data points directly as input. Then, using a tournament selection process, see Ref.~\cite{Bogdanos:2009ib} for more details, the best-fitting functions in each generation are selected and the two stochastic operations of crossover and mutation are used. To guarantee convergence, the GA operation is repeated thousands of times and with various random seeds, to explore properly the functional space. The final output of the code is a set of  functions of $H(z)$ and $d_A(z)$ that describes the Hubble rate and the angular diameter distance respectively.

The error estimates of the reconstructed function are obtained via the path integral approach, originally implemented in Refs.~\cite{Nesseris:2012tt,Nesseris:2013bia}. This approach consists of having an analytical estimate of the error of the reconstructed quantity by computing a path integral over all possible functions around the best fit GA that may contribute to the likelihood, and it has been shown that this can be performed whether the data points are correlated or uncorrelated. This error reconstruction method has been exhaustively examined and tested against a bootstrap Monte Carlo by Ref.~\cite{Nesseris:2012tt}. More explicitly, given a reconstructed function $f(x)$ from the GA, the path integral approach of Ref.~\cite{Nesseris:2012tt} gives us the $1\sigma$ error $\delta f(x)$. This can be also compared to error propagation if one assumes that the error in a quantity is taken as $\sigma_f=f’(p) \delta p$, where $p$ would represent a parameter. We have extensively compared our approach finding that this assumption is appropriate for the data set used here. 

Finally, the reason we need the GA is that traditional inference approaches, such as Markov chain Monte Carlo simulations require a specific model for the expansion history, e.g. $\Lambda$CDM, to fit the data and the choice of this model will clearly bias the results or even miss critical features of the data. The main advantage of the GA is that, given some data, we allow the algorithm to determine the best fit model, thus we can remain theory agnostic, as we do not have to assume a dark energy model.

\section{Results \label{sec:results}}
In this section we present our GA fits to the simulated data and the corresponding consistency test obtained from our reconstructions on $H(z)$ and $d_A(z)$. We want to stress that the aim of this work is not simply to rule out LTB void models as contenders of dark energy models, which has already been done so, but to present a complementary consistency to test of the CP in a non-trivial but still interesting setting.

In Fig.~\ref{fig:FS} we show our new null test $\mathcal{O}(z)$ given by Eq.~(\ref{eq:Otest}) and the $C(z)$ test given by Eq.~(\ref{eq:cp}), both of which can be used to find deviations from the CP through our reconstructions on the Hubble rate $H(z)$ and the angular diameter distance $d_A(z)$. Our reconstructed functions start at $z=0.1$ which is the redshift of our first mock data point. In both cases the black solid line and the gray region correspond to the GA best fit and its $1\sigma$ error respectively. The black dashed-line represents the flat \lcdm model and we see that our reconstructions recover well the null hypothesis of the FLRW metric used to make the mocks. Recall that the \lcdm curves are theoretical and so are precisely at $\mathcal{O}(z)=1$ and $C(z)=0$. The colored lines represent the four different theoretical LTB models, concretely defined in Table~\ref{tab:LTB}. 

In the left panel of Fig.~\ref{fig:FS} we show our $\mathcal{O}(z)$ test and as can be seen, it is a good discriminator of all LTB models at high and intermediate redshifts, i.e. $z\sim 0.6$ and beyond, as the errors remain consistently low at all redshifts. On the other hand, in the right panel of Fig.~\ref{fig:FS} we present the $C(z)$ test which can discriminate the LTB model from \lcdm at intermediate redshifts $0.5<z<1.5$ but at low redshifts ($z<0.3$) and high redshifts above $z\sim 1.5$ the values of $C(z)$ of the fiducial LTB models asymptote to zero, thus being dominated by the errors. 

Therefore, by comparing both panels we may infer that our test manages to detect deviations from the CP particularly well and consistently at middle to high redshifts, when the traditional $C(z)$ test does not perform equally well. Hence, our null test presented serves as a complementary consistency check of the CP and is especially useful at high redshifts.

\section{Conclusions \label{sec:conclusions}}
In summary, we have presented a new consistency test of the Copernican principle, which is complementary to the curvature test of Ref.~\cite{Clarkson:2007pz}. In particular, we used the Noether's theorem approach in order to obtain a conserved quantity that can be written in terms of the Hubble rate $H(z)$ and the comoving distance $D(z)$.

In order to forecast how well our new test, given by Eq.~\eqref{eq:Otest}, can constrain deviations from the Copernican principle at large scales, we created mock datasets based on specifications of the DESI survey and using the \lcdm model for the fiducial cosmology, for a variety of different profiles. This approach allows us to quantify any deviations using forecast mock data and plausible scenarios.

Then, to reconstruct the $\mathcal{O}(z)$ statistic given by Eq.~\eqref{eq:Otest} from the mock data, we preferred to use the machine learning approach, namely the GA, as this will allow us to obtain non-parametric and theory agnostic reconstructions of the data, in the form of $H(z)$ and $D(z)$, that we can in turn use to reconstruct  $\mathcal{O}(z)$. With the same functions we also reconstructed the $C(z)$ function of Ref.~\cite{Clarkson:2007pz} given by Eq.~\eqref{eq:cp2}. 

Following this approach, we find that the GA with the $\mathcal{O}(z)$ statistic can correctly predict the underlying fiducial cosmology at all redshifts covered by the data, as seen in the left panel of Fig.~\ref{fig:FS} and can easily rule out several LTB scenarios at confidence of ~$\gtrsim 3\sigma$ at middle to high redshifts ($z>0.5$). On the other hand, the $C(z)$ statistic, while it successfully rules out the same LTB profiles at small redshifts at a confidence of $\sim 8 \sigma$ at intermediate redshifts ($0.5<z<1.0$), it does not fare equally well at higher redshifts ($z>1.5$) as the errors become larger and the value of $C(z)$ asymptotes to zero, thus diminishing its predictive power. 

To conclude, we find that the $\mathcal{O}(z)$ test provides complementary to other tests, information on possible deviations of homogeneity at different redshift regimes and can help test one of the fundamental assumptions of the standard cosmological model at high redshifts, something which is the goal of several current and upcoming surveys in the coming years.

\section*{Acknowledgements}
The authors acknowledge support from the Research Project No. PGC2018-094773-B-C32 and the Centro de Excelencia Severo Ochoa Program No. SEV-2016-0597. S.~N. also acknowledges support from the Ram\'{o}n y Cajal program through Grant No. RYC-2014-15843.\\

{\bf Numerical Analysis Files}: The Genetic Algorithm code used by the authors in the analysis of the paper can be found at  \href{https://github.com/RubenArjona}{https://github.com/RubenArjona}.\\

\bibliography{cp}

\end{document}